# Unconventional slowing down of electronic recovery in photoexcited charge-ordered La$_{1/3}$Sr$_{2/3}$FeO$_3$

Yi Zhu[1], Jason Hoffman[2], Clare E. Rowland[3,4], Hyowon Park[2,5], Donald A. Walko[1], John W. Freeland[1], Philip J. Ryan[1,6], Richard D. Schaller[3,4], Anand Bhattacharya[2,3*], Haidan Wen[1*]

1. Advanced Photon Source, Argonne National Laboratory, Argonne, Illinois 60439, USA
2. Materials Science Division, Argonne National Laboratory, Argonne, Illinois 60439, USA
3. Center for Nanoscale Materials, Argonne National Laboratory, Argonne, Illinois 60439, USA
4. Department of Chemistry, Northwestern University, Evanston, Illinois 60208, USA
5. Department of Physics, University of Illinois at Chicago, Chicago, Illinois 60607, USA
6. School of Physical Sciences, Dublin City University, Dublin 9, Ireland

* Correspondence and requests for materials should be addressed to H.W. and A. B. (email: wen@aps.anl.gov, anand@anl.gov)

**Ordered electronic phases are intimately related to emerging phenomena such as high T$_c$ superconductivity and colossal magnetoresistance. The coupling of electronic charge with other degrees of freedom such as lattice and spin are of central interest in correlated systems. Their correlations have been intensively studied from femtosecond to picosecond time scales, while the dynamics of ordered electronic phases beyond nanoseconds are usually assumed to follow a trivia thermally driven recovery. Here, we report an unusual slowing down of the recovery of an electronic phase across a first-order phase transition, far beyond thermal relaxation time. Following optical excitation, the recovery time of both transient optical reflectivity and x-ray diffraction intensity from a charge-ordered superstructure in a**

**La$_{1/3}$Sr$_{2/3}$FeO$_3$ thin film increases by orders of magnitude longer than the independently measured lattice cooling time when the sample temperature approaches the phase transition temperature. The combined experimental and theoretical investigations show that the slowing down of electronic recovery corresponds to the pseudo-critical dynamics that originates from magnetic interactions close to a weakly first-order phase transition. This extraordinary long electronic recovery time exemplifies an interplay of ordered electronic phases with magnetism beyond thermal processes in correlated systems.**

The correlation of electronic, spin and structural degrees of freedom in correlated materials have been the basis of emergent phenomena such as high T$_c$ superconductivity, metal-to-insulator phase transitions, and colossal magnetoresistance[1–3]. Strong correlations amongst these degrees of freedom hold the promise for engineering material properties using targeted optical excitation to create hidden phases that does not exist in thermal equilibrium. These hidden phases can live as short as a few picoseconds such as transient superconductivity in cuprates[4,5], and can be long-lived metastable states such as those in photo-excited low-temperature manganites[6] and dichalcogenides[7]. Understanding and possibly controlling how driven quantum system achieves equilibrium is of the central interest in nonequilibrium physics[8] and is essential for elucidating the lifetime of emergent photoinduced physical phenomena at these different time scales.

The charge-ordered state, where patterns of charge density spontaneously emerge below a critical temperature T$_c$, is one of the most interesting collective electronic phases and plays a critical role in determining the properties of many correlated materials[9]. Thanks to the distinct dynamics associated with different interaction mechanisms, complex interactions amongst multiple degrees of freedom such as lattice distortion and spin ordering can be effectively disentangled on ultrafast time scales, motivating studies of femtosecond-to-picosecond (fs-ps) responses[10–14]. Far below $T_c$, charge order (CO) can be quenched by ultrafast optical excitation and typically recovers on ps time scales, during which its relation with nonequilibrium structural distortions has been experimentally studied[15–18]. When the system temperature is close to $T_c$, the initiation and recovery of photoinduced changes are drastically different[10]. For example, a critical slowing down of charge density waves on ps time scales was observed in cuprates[19], molybdenum oxides[20,21], and chalcogenides[11,22]. The recovery of charge ordering in layered organic salts can approach nanosecond (ns) time scale due to electronic instabilities[23]. On longer time scales after the charge, spin, and lattice degrees of freedom have had sufficient time to exchange energy to reach the same temperature, the evolution of electronic properties usually follows the cooling of the system through thermal exchange with the environment. Although the recovery of other degrees of freedom such as magnetism can be slower than the thermal recovery, electronic recovery does not necessarily follow glass-like recovery of magnetic ordering[24]. On these long time scales beyond thermal recovery, electronic slowing down has not been studied, and its microscopic mechanisms and the relation to other degrees of freedom are not clear. Understanding the role of correlations on these mesoscopic time scales is essential to extending the lifetime of exotic electronic phases beyond thermal cooling of the system.

Here, we report on an unusually slow recovery of a collective electronic phase that becomes significantly longer than the lattice cooling time near $T_c$ in photoexcited $La_{1/3}Sr_{2/3}FeO_3$ (LSFO) thin films. Using time-resolved optical spectroscopy and x-ray diffraction, we directly track three quantities in the time domain: the optical reflectivity, the superlattice x-ray diffraction peaks related to the order parameter of CO, and the lattice constant of LSFO [Fig. 1(a)]. When the sample temperature increases towards $T_c$, we observe concurrent slowing down of the recovery of both transient optical reflectivity and CO superlattice diffraction peak intensity, well beyond the few-ns thermal recovery of the out-of-plane lattice constant as independently characterized. Moreover, the x-ray diffraction measurements reveal important structural information: no significant change of the CO domain size is observed, which rules out a mesoscopic nucleation and growth scenario and suggests a microscopic origin for the slowing down. By density functional theory plus U (DFT + U) calculations, we find that the magnetic-exchange-driven phase transition is weakly first order, in which phase separation of two competing CO configurations occurs close to $T_c$. The recovery of CO following the pathways along the temperature-dependent potential energy surfaces to the ground state qualitatively explains the observed slowdown. The scaling of the time constants as a function of temperature is ascribed to pseudo-critical dynamics in a weakly first-order transition[25], driven by magnetic exchange interactions.

The perovskite oxide $La_{1/3}Sr_{2/3}FeO_3$ is a prototypical material in which the superexchange interaction between Fe sites mediated via oxygen ions drives a first-order metal-to-insulator (MIT) phase transition[26–28]. The MIT, paramagnetic to antiferromagnetic (AFM) transition, and CO transition are concurrent below a transition temperature ($T_c$) of ~200 K. A 70 nm LSFO thin film sample was grown by ozone-assisted molecular beam epitaxy (MBE) on (111) $SrTiO_3$ (STO) substrates. The $T_c$ of LSFO film was experimentally identified as the temperature at which charge ordering emerges [Fig. 1(b)]. Upon cooling below $T_c$, charge disproportionation between the Fe sites leads to charge ordered planes of $Fe^{5+}Fe^{3+}Fe^{3+}$ stacked along the [111] direction in the pseudo-cubic representation[26], accompanied by a sharp hysteretic rise in resistivity [Fig.1(b)]. While the reported valence states of Fe ions vary from 3+ to 5+ [27,28], an accompanying periodic structural distortion gives rise to CO superlattice ($n\pm1/3$, $n\pm1/3$, $n\pm1/3$; $n$ is a positive integer) peaks in x-ray diffraction measurements[28,29]. Unlike the organic Mott insulators[23], the formation of CO in LSFO is a first-order phase transition driven by magnetic interactions[26–28].

The transient electronic and structural dynamics were measured by time-resolved optical reflectivity and x-ray diffraction at the Center of Nanoscale Materials and the Advanced Photon Source at Argonne National Laboratory (see Methods). Calibrated sample temperatures are used throughout this paper (see Supplemental Materials). In both optical and x-ray measurements, the photon energy of the pump laser is above the optical absorption edge of the LSFO thin film (2.2 eV)[30]. The penetration depth of the pump laser is ~30 nm[30], comparable to the optical probing depth but smaller than the x-ray probing depth of 70 nm, the thickness of the film. The mismatch between pump and probe penetration depth in the x-ray measurements is not a concern on the time scale of interest here (ns to μs), which is well beyond the thermalization time of the film[32].

The changes of optical reflectivity measured at 1.1 μm following optical excitation are shown at various temperatures in Fig. 2(a). The recovery process can be fit by an exponential decay function $\Delta R(t) \sim A_1 \exp(-\frac{t}{t_{fast}}) + A_2 \exp(-\frac{t}{t_{slow}})$ with time constants $t_{fast}$ and $t_{slow}$. Far below $T_c$, e.g. at T= 176 K, the fitting yields $t_{fast} = t_{slow} \sim$ 2.3 ns, thus effectively a single exponential decay is sufficient to describe the recovery dynamics. However, as the sample temperature approaches $T_c$, $t_{slow}$ increases orders of magnitude while $t_{fast}$ does not change significantly, as summarized in Fig. 2(b). This observation supports that $t_{slow}$ is related to a non-thermal recovery mechanism while $t_{fast}$ is consistent with thermal recovery as discussed later. The measurement above $T_c$ shows the similar several-ns recovery as observed far below $T_c$, since the recovery of optical reflectivity above $T_c$ is mainly driven by the thermal recovery. The decrease in reflectivity at 1.1 μm upon optical excitation and the slowing down of the recovery are universal across the probed optical spectrum of 0.9 to 1.3 μm (Supplementary Materials). The change of optical reflectivity is consistent with the spectral weight transfer of the optical conductivity from around 1 eV to low energy (<0.5 eV) as the film temperature increases[31]. This spectral range was ascribed to the transition from O 2p to $Fe^{3+}/Fe^{5+}$ $e_{g\uparrow}$ states and is closely related to the macroscopic Drude response across the metal-to-insulator phase transition[31].

In order to understand the slowing down of the recovery of the electronic degree of freedom, we studied the recovery of structural distortion as a result of CO as well as the lattice constant by monitoring the (4/3,4/3,4/3) and (2,2,2) x-ray diffraction peaks, respectively. Since the lattice remains rhombohedral (R3c space group) across $T_c$, the time-dependent lattice constant provides an independent measurement of the film temperature during cooling of the film. At T = 121 K, the CO recovery follows the lattice dynamics. Radial scans of both (4/3,4/3,4/3) and (2,2,2) peaks were measured as a function of delay between the pump laser and probe x-ray pulses [Fig. 3(a)]. The intensity of the superstructure peak decreases as CO melts, while the peak center shifts to lower *HKL,* indicating a superlattice expansion. The expansion strain of 0.04 % measured at 100 ps corresponds to the film temperature increase of 35 K, calculated using the thermal expansion coefficient of $1.16 \times 10^{-5}$ / K [33]. No discernible change in the CO peak width was observed, indicating a nearly constant CO domain size upon photo-excitation. In addition, we found the photoinduced strain measured by the shifts of the (4/3,4/3,4/3) and (2,2,2) peaks relaxed at the same rate [Fig. 3(b)]. The intensity of the superstructure peak (not shown) and transient expansion of the lattice can both be fit by a single exponential function with recovery time constants, $\tau_{CO}$ = 3.1 ns and $\tau_{lattice}$ = 2.8 ns, in good agreement with the recovery time constant of optical reflectivity at a temperature far from the CO phase transition. These observations show that, at temperatures far below $T_c$, the recovery of the electronic degree of freedom is determined by the cooling rate of the thin film. Closer to $T_c$, at T = 196 K, Fig. 3(c) shows the radial scans of the CO peak before and after the laser excitation. The CO peak intensity is suppressed upon laser excitation, while no discernable change to the peak width was observed. Comparing with the CO intensity recovery measured at 121 K in Fig. 3(d), the recovery time ($\tau_{CO}$) of the CO peaks increases two orders of magnitude from 3.5 ns at 121 K to 263 ns at 196 K. Meanwhile, the recovery time of (2,2,2) peak only increases by a factor of 1.5. The film temperature relaxation is well-described by one-dimensional thermal transport to the substrate[32,34]. At tens of ns, the relaxation can be modeled by an exponential decay plus an offset. A careful evaluation of time-dependent film temperature shows that the thermal recovery pathway cannot explain the observed slowing down (Supplementary Materials, Section 4).

To quantify the dynamical scaling of slowing down, we plot the recovery time constants of the CO and lattice diffraction peaks as a function of the local film temperatures in Fig. 3(e). Approaching $T_c$, $\tau_{CO}$ starts to increase dramatically by orders of magnitude, while $\tau_{lattice}$ remains below 7 ns across the CO phase transition. The deviation of the CO superlattice dynamics from that of the lattice peak clearly indicates the non-thermal nature of the recovery of the CO phase. Under similar pump laser fluence, the concurrent slowing down of the recovery of both the optical reflectivity and the CO superlattice peak around 190 K suggests that the dielectric constant at 1.1 μm are also related to the charge ordering, although the probe photon energy of 1.13 eV is much higher than the CO-induced energy gap of 0.13 eV[31]. This observation indicates the macroscopic optical properties across a wide spectrum range in LSFO are highly correlated with the formation of charge ordering on longer-than-thermal recovery time scales. Measurements at other pump fluences are consistent with this observation (Supplemental Materials). The time constant of the slowing down as a function of temperature are fit to a power law of $\tau = \tau_0 (1-T/T_c)^{-\Delta}$ where $T_c$ = 200 K and $\Delta$ is the scaling exponent and $\tau_0$ is a constant[25]. The best fit yields $\Delta = 1.25 \pm 0.10$, shown by the magenta curve in Fig. 3(e). The time constant of optical reflectivity recovery is fit by the same function and shown as the pink curve in Fig. 2(b), with $\Delta = 1.06 \pm 0.16$. In conventional second-order critical phenomena, the measured scaling exponent $\Delta$ is equivalent to $vz$ ($v$ and $z$ are two critical indices[35]) and approximately agrees with 1.3 and 1.37 for three-dimensional (3D) Ising[36] and Heisenberg[37] models respectively, and differs from 2.16 for a two-dimensional models[38,39]. However, we point out that the phase transition in LSFO is first order like, because a latent heat is presented at the transition in bulk samples[40]. In addition, our electron transport measurements [Fig. 1(b)] and the antiferromagnetic order parameter reported previously[29] are weakly hysteretic as a function of temperature. Thus, our measurements show a pseudo-critical phenomenon near a weakly first-order phase transition[25], rather than conventional second-order critical phenomena[35]. We also note that the attempted fit for the nucleation and growth model does not agree with measurements (Supplemental Materials). Therefore, the observed scaling cannot be explained by the reduction of nucleation rate of CO domains as system temperature approaches $T_c$, consistent with no discernible changes of coherent length of CO domains upon optical excitation.

To further understand the origin of the slowing down, we calculated the total energies of LSFO by adopting the first−principles DFT + U method, which illustrates the recovery pathways of coupled degrees of freedom during a first-order phase transition. First, we performed the structural relaxation using U = 5 eV and J = 1 eV. We found that the ground state of LSFO is charge ordered with an associated structural distortion as a result of antiferromagnetic ordering, consistent with the previous DFT + U calculation[41]. The resulting structural distortion is characterized by the average Fe-O bond-length difference $\delta_a$=0.06 Å (inset, Fig. 1b) between different Fe sites, with the averaged AFM moment of 3.7 $\mu_B$. Since DFT+U is a zero-temperature theory, we simulate the temperature effect by varying the interaction parameters (U and J) to tune the resulting magnetic moment, which is controlled by the sample temperature experimentally. Without any magnetic interactions as will be realized in very high temperature, i.e., U = J = 0, the energy surface shows no CO and accompanying structural distortion. The calculation thus agrees with the magnetic-driven CO in LSFO[27]. The structural pathway for the energy surface along $\delta_a$ is determined by interpolating the CO structure relaxed using U = 5 eV and J = 1 eV and the non-CO structure relaxed using U = J = 0 eV. We then vary U values from U = 5 eV for the low-temperature CO state to U = 0 eV for the high temperature state without CO, while fixing the U/J ratio to 5. This is used to study the qualitative features of the energy landscapes between two local minima by simulating an increase of sample temperature. The structural pathway along $\delta_a$ between two local minima is fixed while U values are changed and we explore the energy landscapes qualitatively due to the change of temperatures. While the resulting magnetic moments and energy landscapes can be changed by tuning the U values, the calculated AFM moments for reasonable U values (3-5 eV) are on the order of 3 $\mu_B$, comparable with the experimental values[42,43]. At T<<$T_c$, our calculations show that the large AFM spin exchange energy between $Fe^{3+}$ ions dominates in energetics and gives rise to one energy minimum state at $\delta_a$ = 0.06 Å, accounting for $Fe^{5+}$-$Fe^{3+}$-$Fe^{3+}$ order or small-large-large oxygen octahedrons, as illustrated by diamonds in Fig. 4(a). As the AFM moment was reduced as the sample temperature increases, we discovered that a metastable state starts to emerge at $\delta_a$ = -0.06 Å, with $Fe^{4+}$-$Fe^{4+}$-$Fe^{3+}$ ordering or small-small-large oxygen octahedrons. By reducing the value of U to 3.7 eV, we obtained the magnetic moment 3.3 $\mu_B$ at which the energy of this metastable state at $\delta_a$ = -0.06 Å is further lowered and becomes degenerate with the state at $\delta_a$ = 0.06 Å, giving rise to the coexistence of two competing CO states with nonzero energy barrier. While the magnetic moments and relative energies of two CO states may sensitively depend on the value of U, the position of the two CO states, i.e., the values of $\delta_a$ at two local energy minima, are not sensitive to U.

The DFT + U calculation shows the energetics of the system is driven by magnetic exchange. Comparing with a typical first-order [Fig. 4(b)] or second-order phase transition, the energy surfaces exhibit unique characteristics that governs the electronic dynamics. First, the energy barrier $E_b$ is small but persist with non-zero AFM moments, which is consistent with the signature of a weakly first order instead of a typical second-order phase transition. Second, the existence of two nearly degenerate energy states and the rapid quench following the heat pulse provides the opportunity for the system to enter the regime of spinodal decomposition. In this scenario, following the initial laser excitation, the system is 'quenched' with $\delta_a = 0$ (disordered state) from temperatures above $T_c$ due to relatively fast cooling of the film temperature on the ns timescale. At this point, the system is in the unstable 'spinodal' regime where the second derivative of the free energy is less than zero. There is no energy barrier to the formation of the CO states, and the kinetics is limited only by diffusion[44]. Furthermore, this process slows down in the vicinity of spinodal points where second derivative of the free energy is zero [Fig. 4(c)], and the diffusion constant becomes vanishingly small[25]. During this process, the domain size does not change, consistent with our observation. At later stages beyond our measurement time window, processes similar to Ostwald ripening can occur and would increase the coherence length[44]. A direct observation of the associated kinetics of spinodal decomposition needs time-resolved resonant x-ray diffraction microscopy with sufficient spatiotemporal resolution which is beyond the scope of this work. At low temperatures, as schematically shown in Fig. 4(c), the deep potential well at $T \ll T_c$, i.e. with strong restoring force, gives rise to a fast recovery from the excited state following a well-defined recovery pathway.

In summary, we observed a slowing down of the recovery of an electronic phase longer than the thermally driven processes in a photo-excited LSFO thin film. The multimodal probes via transient optical reflectivity and CO superlattice diffraction peak allow direct correlation between optical property and the long-range electronic ordering. The electronic recovery is significantly different from the relaxation of the average lattice parameter, providing decisive evidence of non-thermal recovery of the electronic phase. First principle DFT + U calculations elucidate a microscopic picture of magnetic-interaction driven slowing down and suggest a pseudo-critical phenomenon close to a weakly first-order phase transition. Our combined experimental and theoretical investigation provides experimental verification and mechanistic insight on an unconventional critical behavior and the interplay of multiple degrees of freedom on unusually long time scales at an electronic phase transition. This work opens up opportunities of controlling nonequilibrium processes in correlated systems beyond time scales for thermal recovery.

**Methods**

***Sample preparation and characterization.*** Epitaxial $La_{1/3}Sr_{2/3}FeO_3$ (LSFO) thin films were grown using ozone-assisted molecular beam epitaxy (MBE) on (111)-oriented $SrTiO_3$ (STO) substrates. Prior to growth, trichloroethylene was used to remove organic contaminants from the substrate surface. The co-deposited elemental materials La, Sr, and Fe were evaporated from effusion cells under an ozone environment with partial pressure of $3\times10^{-6}$ mbar, with the substrate temperature maintained at 680° C. The evaporation rates were determined for each material from a quartz crystal thickness monitor that was calibrated to within 2% from Rutherford backscattering measurements. The film thickness and surface symmetry were monitored in real time from reflection high-energy electron diffraction (RHEED) intensity oscillations. Brief anneal periods of ~30 seconds followed the completion of each unit cell layer. After the LSFO deposition, the samples were cooled down to room temperature in an environment of $3\times10^{-6}$ mbar of $O_3$. The 70-nm thick LSFO sample was characterized by static x-ray diffraction and the results are shown in Fig. S1. A radial x-ray diffraction scan along the [111] direction at 121 K shows the charge order (CO) (4/3, 4/3, 4/3) peak and the STO and LSFO lattice (1, 1, 1) peak. The integrated intensity of the CO peak was measured as a function of the sample temperature, showing the CO emerges at ~200 K, and four-point probe measurement of the sheet resistance shows a hysteresis loop close to the transition temperature, shown in Fig. 1 (b) of the main text.

***Experimental setup.*** In the ultrafast optical pump-probe experiment shown schematically in Fig. S2 (a), an optical parametric amplifier (OPA) was pumped by a femtosecond Ti:Al$_2$O$_3$ laser at 1 kHz repetition rate. The output wavelength of the OPA was doubled to $\lambda = 420$ nm to excite the LSFO sample, with a pulse duration of 100 fs. A Nd:YAG laser was electronically synchronized with the femtosecond laser to perform asynchronous optical sampling measurements. The output of the YAG laser was focused into a sapphire plate to generate white light with wavelength from 900 nm to 1300 nm. The probing white light was re-focused to be smaller than and spatially overlapped with the pump laser beam on the LSFO sample surface. The reflected white light was analyzed by a spectrometer. The sample temperature was controlled from 77 K to room temperature. In the time-resolved hard x-ray diffraction experiment [Fig. S2 (b)], the pump laser pulse was derived from the third harmonic generation (THG) of a high repetition rate (54 kHz) Nd:YAG laser, with 355 nm central wavelength and ~10 ps pulse duration. Use of a high-repetition-rate laser is essential to achieve high signal-to-noise ratio for probing time-resolved CO x-ray diffraction. X-ray pulses at 12 keV photon energy and ~100 ps pulse duration were focused by Kirkpatrick-Baez mirrors to a beam size of ~50 μm, smaller than the focused pump laser beam size of ~220 μm. An area detector (Pilatus100K, DECTRIS Ltd.) gated at 54 kHz was used to detect the diffraction intensity. The LSFO sample was mounted on a six-circle diffractometer (Huber GmbH.) in a cryostat with temperature control from 30 K to 300 K.

***DFT calculation.*** The DFT + U calculations were performed using the Vienna ab-initio simulation package (VASP). We adopted the generalized gradient approximation (GGA) exchange-correlation functional for simulations using an energy cutoff of 600 eV and a k-point mesh of 8×8×2 aligning the z-axis toward the [111] direction. We use the supercell of LSFO to accommodate both the antiferromagnetic spin configuration and the octahedral expansion or collapse. The supercell elongated along the [111] direction contains 2 La atoms, 4 Sr atoms, 6 Fe atoms, and 18 O atoms allowing the expansion or the collapse of 6 $FeO_6$ octahedral volumes. The convergence of the total energy calculation was reached if the consecutive energy difference was within $10^{-4}$ eV and the atomic forces of all ions were required to be smaller than 0.01eV/Å for ionic relaxations. The "+U" interaction Hamiltonian was adopted using the rotationally invariant form and the DFT Hamiltonian part did not include the spin-exchange splitting (i.e., the DFT energy part was a function of spin un-polarized charge density, and the spin exchange interaction is entirely treated within the correlated Fe d orbitals) aligning with the spirit of DFT+ dynamical mean field theory, as implemented in VASP using the option of "LDAUTYPE=4". The widely used spin-DFT+U implementation, i.e., the spin polarization accounts for both the charge density and the correlated orbitals, often overestimates the spin-exchange interaction in several other systems[45–47]. Especially for this LSFO material, the resulting magnetic moments and energy curves computed using spin-DFT+U were insensitive to the change of U and J parameters since the DFT part alone already accounts for the large portion of the spin-exchange interaction.

**Acknowledgement**

We acknowledge G. Doumy and A. M. March for the assistance of using the high-repetition rate laser at the 7ID-C beamline. Y. Z., J. D. H., A. B. and H. W. acknowledge support of U.S. Department of Energy (DOE), Office of Science, Basic Energy Sciences (BES), Materials Sciences and Engineering Division. H. P. acknowledges support of the start-up funding from the University of Illinois at Chicago and Argonne National Laboratory (by the U.S. Department of Energy, Office of Science program) and the computing resources provided on Blues, a high-performance computing cluster operated by the Laboratory Computing Resource Center at Argonne National Laboratory. H. W. and J. F. acknowledge the support of data analysis by DOE-BES grant No. DE-SC0012375. The use of the Advanced Photon Source and Center for Nanoscale Materials is supported by DOE-BES, under Contract No. DE-AC02-06CH11357.

**Author contributions:** H. W. and A. B designed the project. Y. Z., D. W., J. F., P. R., H. W. did the x-ray measurements. C. R. and R. S. did the optical measurements. J. H. and A. B. made the samples. Y. Z. and H. W. wrote the paper with the contribution from all authors.

**Additional information**

Supplementary Materials is available in the online version of the paper.

Competing financial interests: The authors declare no competing financial interests.

**Figures and Captions:**

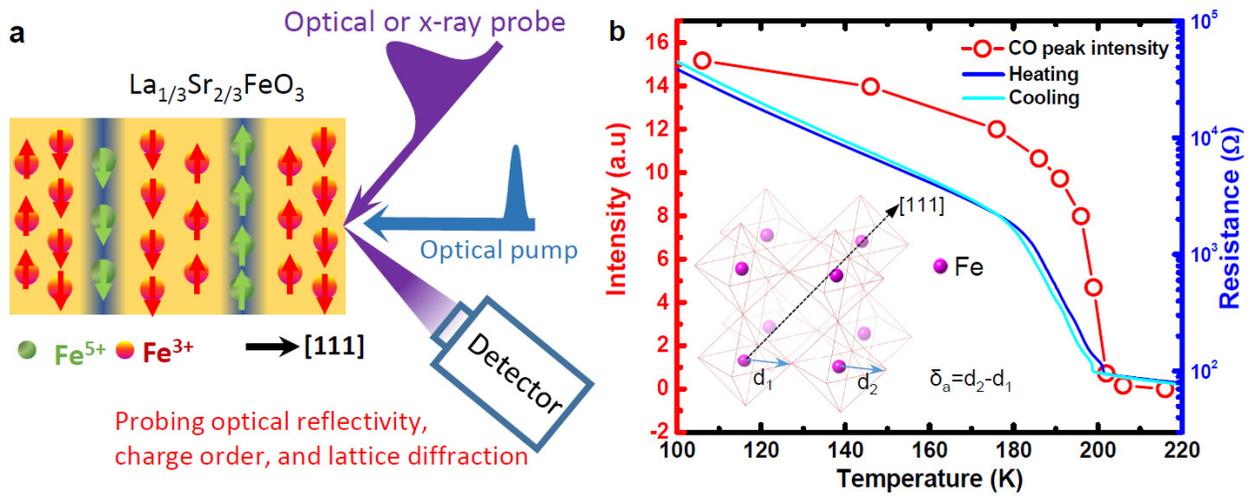

**Figure 1:** (a) Schematic drawing of the LSFO charge and AFM ordering and the setup of time-resolved optical and x-ray diffraction experiments. Below $T_c$, $Fe^{3+}$ and $Fe^{5+}$ ions order to form a periodic charge density distribution along the pseudo cubic [111] direction. Upon above-band-gap optical excitation, the transient optical reflectivity and the time-resolved x-ray diffraction from CO (4/3, 4/3, 4/3) superlattice and (2, 2, 2) lattice was measured. (b) The temperature-dependent CO diffraction intensity (red circle) and the resistance (blue for heating and cyan for cooling) of the sample. The inset shows oxygen octahedrons with the differentially averaged Fe-O length $\delta_a = d_1 - d_2$.

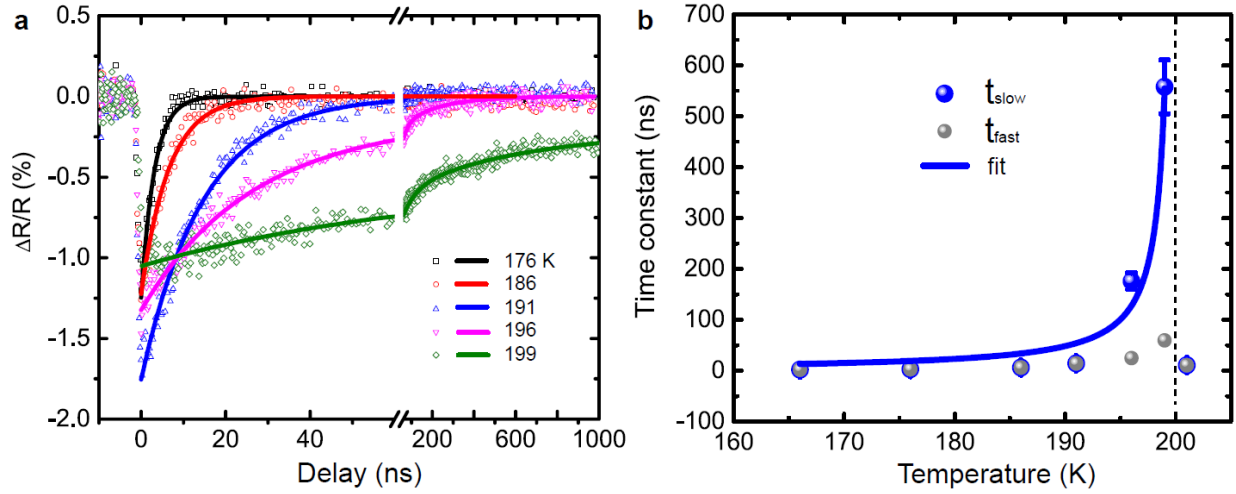

**Figure 2** (a) Transient reflectivity changes probed at the wavelength of 1100 nm at various sample temperatures. Solid lines are fits with the fitting errors showing as error bars in (b). (b) Recovery time constants of the transient reflectivity change as a function of sample temperatures. The solid line is a fit of $t_{slow}$ to the scaling law $\sim (1 - T/Tc)^{-\Delta}$. The absorbed pump laser fluence is 2.9 mJ/cm$^2$.

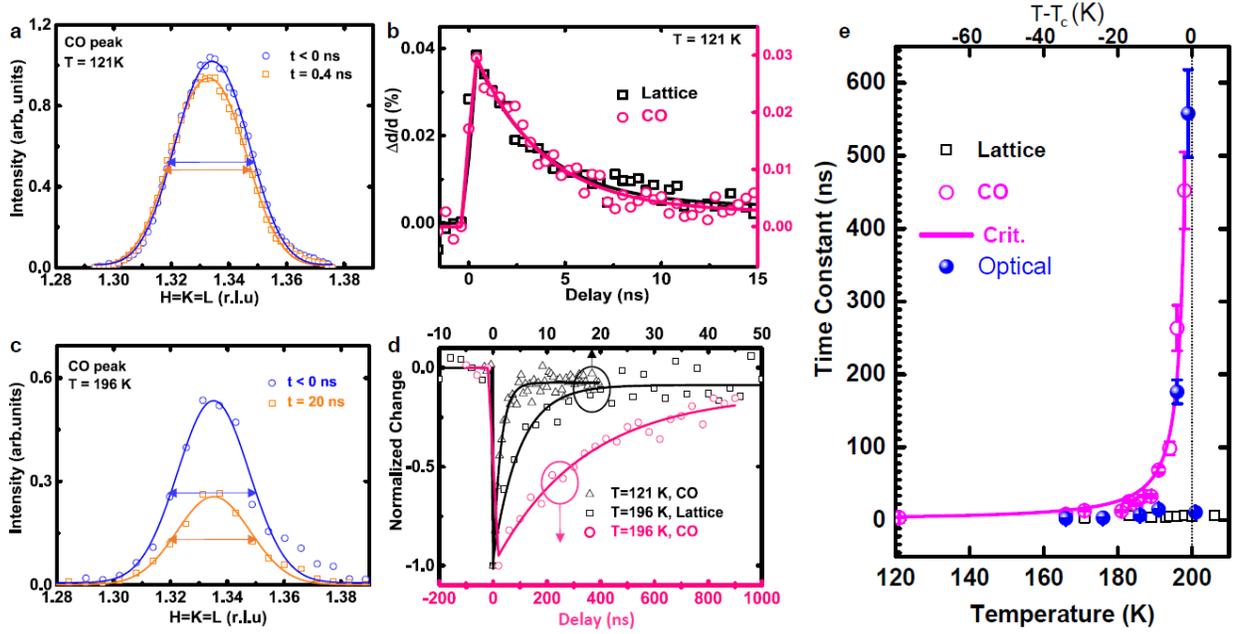

**Figure 3** (a) Radial scans of CO peak measured at 121 K at various delays. The solid lines are Gaussian fits. The double-arrow lines with identical length show that the full width half maximum of the peaks are equal within resolution. The HKL are indexed with respect to the LSFO ground-state lattice at specified temperatures. (b) Out-of-plane strain of lattice (black square) and CO superlattice (pink circle) as a function of the delay, with the fit result shown as solid lines. (c) Radial scans of CO peak measured at 196 K at various delays. (d) CO peak intensity as a function of delay is measured at T = 121 K (black triangles) and 196 K (pink circles). Lattice strain as a function of time measured at T = 196 K (black squares). The color of the curve matches the color of the corresponding x axis. (e) Recovery time constants of CO intensity $\tau_{CO}$ and lattice peak shift $\tau_{lattice}$ measured as a function of sample temperature at an absorbed pump laser fluence of 2.9 mJ/cm$^2$. The solid line shows the fit of x-ray data based on the power law $(1-T/T_c)^{-\Delta}$. The dotted line indicates the nominal transition temperature. The blue solid balls show the recovery time constant of optical reflectivity same as in Fig. 2(b).

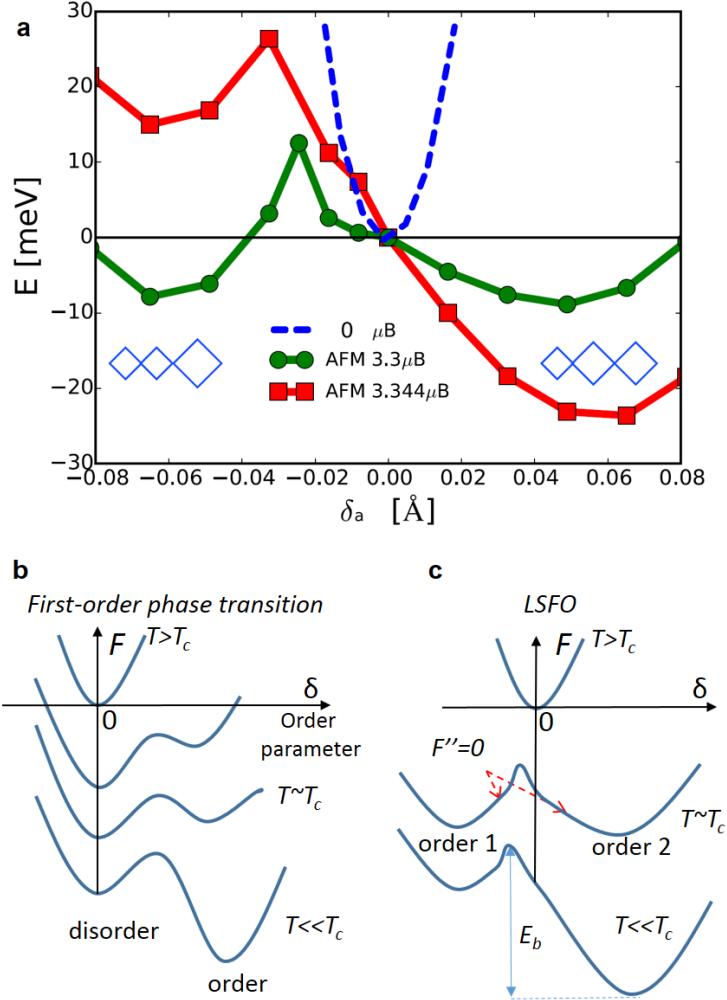

**Figure 4** (a) The total energy curve calculated using DFT+U as a function of the Fe-O bond length difference $\delta_a$ [see inset Fig. 1(b)] along [111] axis at various Bohr magnetons. Different magnetic moments are obtained by changing U and J values in DFT+U while fixing the U/J ratio. The AFM ordering is the ground state in LSFO within DFT+U. The arrays of blue diamonds represent the ordering of the small and large oxygen octahedrons around $\delta_a = \pm 0.06$ Å. (b) The schematic energy potential surfaces for the first-order phase transition at varies temperatures; (c) The energy potential surface for LSFO shows the degenerate charge ordered states around the transition temperature. $E_b$ is the energy barrier between two competing charge-ordered states.

# Supplemental Materials for "Unconventional slowing down of electronic recovery in photoexcited charge-ordered La$_{1/3}$Sr$_{2/3}$FeO$_3$"

Yi Zhu[1], Jason Hoffman[2], Clare E. Rowland[3,4], Hyowon Park[2,5], Donald A. Walko[1], John W. Freeland[1], Philip J. Ryan[1,6], Richard D. Schaller[3,4], Anand Bhattacharya[2,3]*, Haidan Wen[1]*

1. *Advanced Photon Source, Argonne National Laboratory, Argonne, Illinois 60439, USA*
2. *Materials Science Division, Argonne National Laboratory, Argonne, Illinois 60439, USA*
3. *Center for Nanoscale Materials, Argonne National Laboratory, Argonne, Illinois 60439, USA*
4. *Department of Chemistry, Northwestern University, Evanston, Illinois 60208, USA*
5. *Department of Physics, University of Illinois at Chicago, Chicago, 60607, USA*
6. *School of Physical Sciences, Dublin City University, Dublin 9, Ireland*

*wen@aps.anl.gov, anand@anl.gov

## 1. X-diffraction measurements

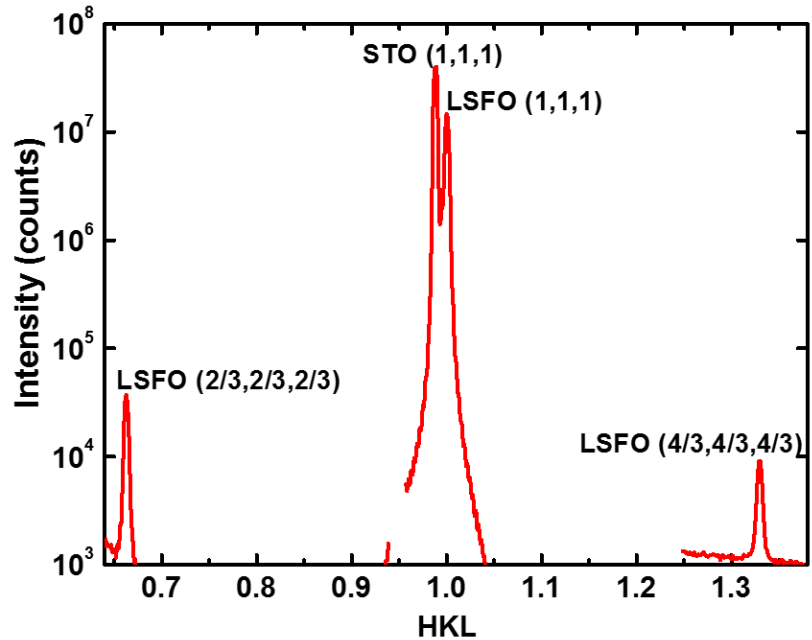

FIG. S1 The radial scan along the out-of-plane [111] direction at T=121 K. The HKL is indexed with respect to the Brag peak of LSFO film of the ground state at the specified sample temperature.

## 2. Experimental setup

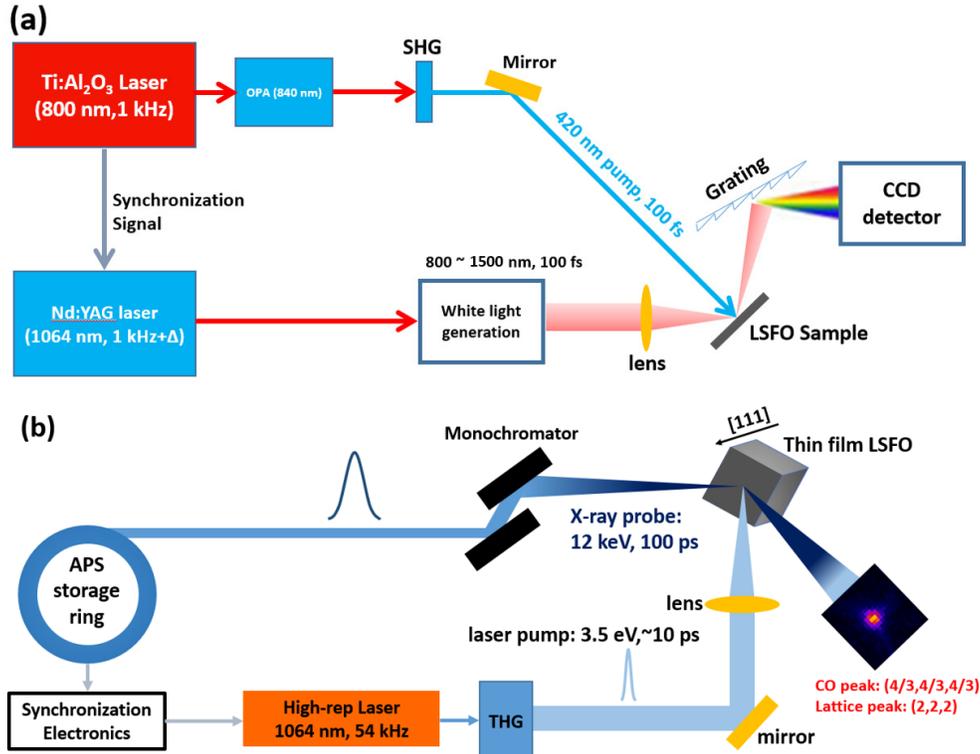

FIG. S2 Experimental setup. (a) Ultrafast optical pump-probe experimental setup. OPA: optical parametric amplifier; SHG: second harmonic generation. (b) Time-resolved hard x-ray diffraction experimental setup. THG: third harmonic generation.

## 3. Transient optical reflectivity measurements

The slowing down of the recovery of optical reflectivity is universal across the probing spectrum close to $T_c$. This is clearly shown in two representative time-dependent spectroscopy plots at 196 K and 199 K, respectively (Fig. S3). In the main text, we select the optical probe wavelength of 1100 nm to represent the slowing down of reflectivity recovery because the probing spectral intensity close to the fundamental wavelength of 1064 nm gives better signal-to-noise ratio (SNR). The SNR is reduced at the probe wavelength far away from 1064 nm since the white light spectrum intensity is weaker. The spectrum abnormality from 1060 to 1070 nm is due to a notch filter used to filter the fundamental wavelength.

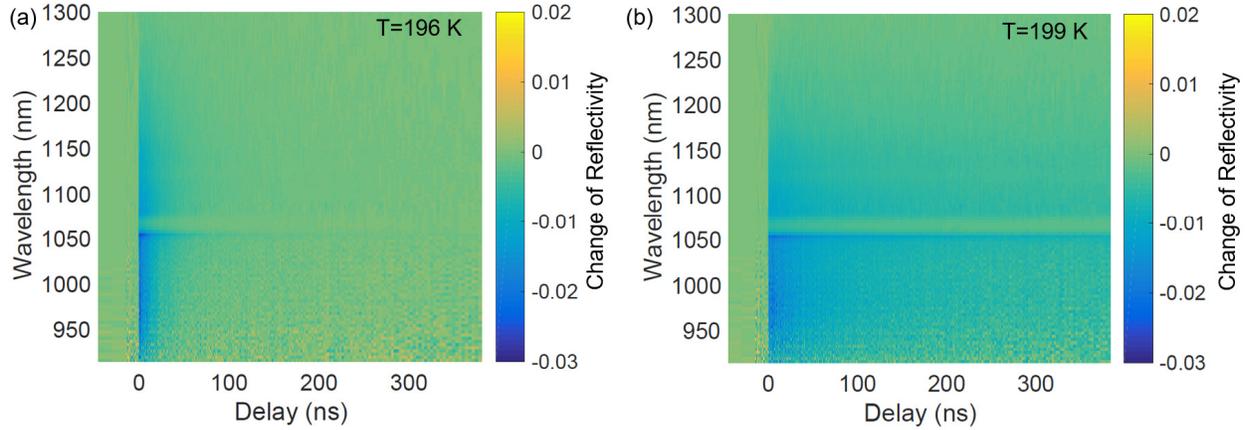

FIG. S3 The time- and spectra- dependent reflectivity change measured at 196 K (a) and 199 K (b). The spectrum gap between 1050-1070 nm is an artifact due to the notch filter for filtering the probe wavelength at 1064 nm used for white-light generation.

### 4. Time-dependent lattice constant

The representative lattice dynamics as measured by the change of (2, 2, 2) Bragg reflection as a function of time at 121 K and with laser fluence of 2.9 mJ/cm$^2$ is shown in Fig. S4. Similar observations are recorded at various temperatures and the recovery of the lattice constant is summarized in FIG. 3e of the main text.

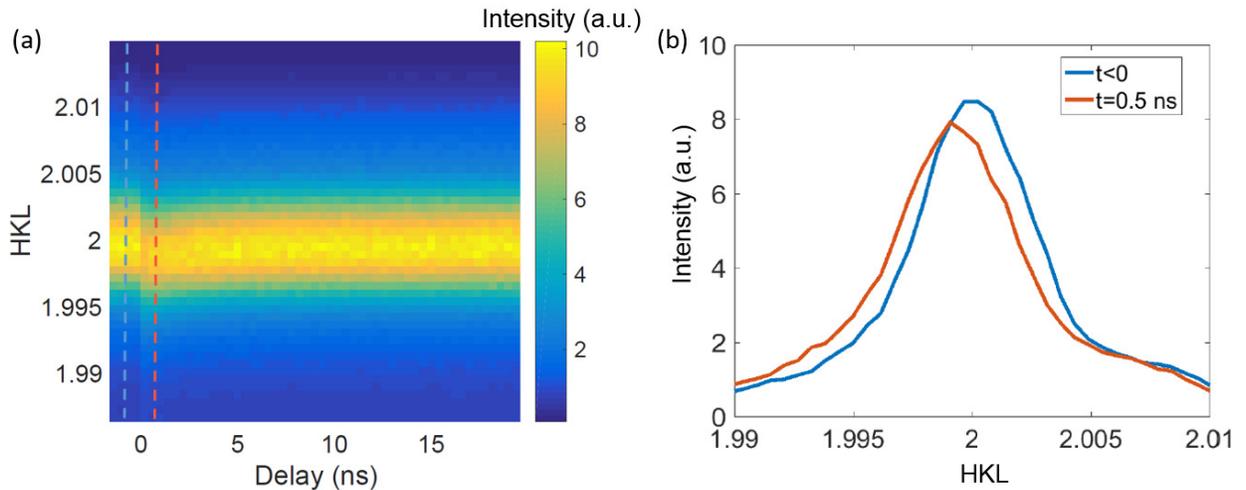

FIG. S4 Lattice (2, 2, 2) dynamics. (a) Radial scans of the lattice (2, 2, 2) peak as a function of delay. (b) The radial scans of the lattice before and after laser excitation as marked by the corresponding color dashed lines in (a).

The cooling of the film is well understood by a one-dimensional thermal transport model and verified by time-resolved x-ray diffraction measurements [e.g., see Sec. III. Ref. [34], Sec. 4 in Ref. [32]]. Using parameters listed in the table of Fig. S5, we can calculate the film temperature by solving the thermal transport equation. The results are shown as solid curve in Fig. S5.

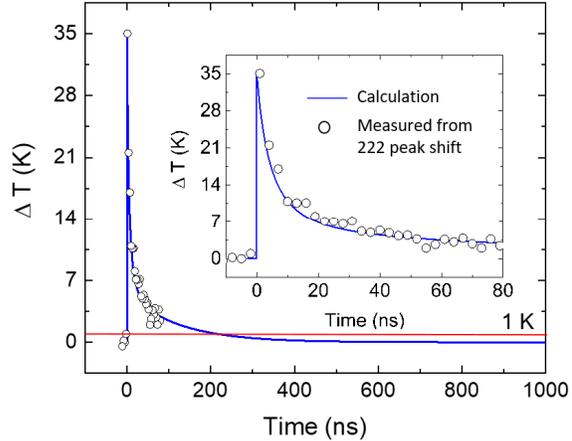

Fig. S5 The calculated (solid line) and measured (open circles) temperature change of the film as a function of time. The maximum of temperature jump ΔT at 100 ps was calculated based on the shift of 222 Bragg peak. The right table shows the parameters used for 1D thermal transport calculation.

| Parameters | La$_{1/3}$Sr$_{2/3}$FeO$_3$ | SrTiO$_3$ |
|---|---|---|
| Thermal conductivity (Wm$^{-1}$K$^{-1}$) | 5 [1] | 10.7 [2] |
| Heat Capacity (J g$^{-1}$ K$^{-1}$) | 0.4 [3] | 0.57 [4] |
| Kapitza Constant (Wcm$^{-2}$K$^{-1}$) | 10000 [5] | |

[1] Phys. Rev. B, **64**, 024421 (2001)
[2] Appl. Phys. Lett. **92**, 191911 (2008).
[3] Approximate by the value of LaFeO$_3$, J. Chem. Thermodynamics **30**, 365 (1998)
[4] Phys. Solid State **51**, 1189 (2009).
[5] Approximate by the value of the oxide interface, Phys. Rev. B, **94**, 180104 (2016)

At tens of ns time scales, this numerical results also can be approximated by an exponential decay function plus an offset. The excellent agreement between the simulation (solid curves) and measurements (open circles) allows us to use the calculated results to reliably predict the sample temperature. For example, the small offset at 200 ns corresponds to the film temperature of 1 K, which is close to a full transient thermal recovery. At this delay, however, the CO diffraction intensity is still far from a full recovery as shown by the pink curve in main Fig. 3(d), indicating a nonthermal origin.

## 5. Steady-state sample temperature calibration and fluence dependent measurement

Due to the high repetition rate of the pump laser (54 kHz) used in the time-resolved x-ray diffraction experiments, the local temperature of the sample under pump laser illumination was higher than the temperature measured by the platinum resistor (PT-100) sensor attached to the sample holder. We calibrated the "steady-state" sample temperature by comparing the CO diffraction intensity with laser illumination to the static temperature-dependent CO intensity measurements, shown in Fig. S6 (a). In this regime below the phase transition temperature of 200 K, the sample temperature is proportional to the incident laser power since the involved latent heat is a small fraction of the total latent heat, which is also a small value for LSFO (see Ref. [40]).

The CO peak intensity measured 1 ns before laser excitation, with the sample holder reading a fixed T = 178 K was used to gauge the local temperature by comparing it with the static temperature-dependent CO intensity. By adjusting the scaling of the horizontal axis on the bottom of Fig. S6 (a), the change of diffraction intensity as a function of absorbed fluence (square) was matched to that as a function of temperature (triangle), so that the bottom (fluence) and top (temperature) axes have a one-to-one mapping. For example, at the absorbed pump laser fluence 2.9 mJ/cm$^2$ indicated by the red arrows, the photoinduced CO peak intensity decrease (-16%) is comparable to that when the sample temperature increases from 178 to 191 K, which means the local sample temperature is 13 K higher than the measured sample holder temperature. The local sample temperature at other pump laser fluencies thus can be calculated accordingly.

The accuracy of the sample temperature calibration is further confirmed by the slowing down of the recovery dynamics measured under different pump laser fluences. Fig. S6 (b) summarizes the recovery time constants as a function of calibrated temperatures, measured by CO diffraction peak and the optical reflectivity probes. For the optical pump-probe experiment at 1 kHz repetition rate, the average pump laser heating effect is negligible. For time-resolved x-ray diffraction experiments, the corresponding heating effect at absorbed pump laser fluences of 2.9, 2.3 and 1.5 mJ/cm² is 13 K, 10.3 K and 6.7 K respectively. After calibrating the temperature, the slowing down starts to occur around 190 K [Fig. S6 (b)], regardless of the pump fluences, indicating that the calibration accurately provided the local film temperature. We estimate the temperature calibration error is <1 K.

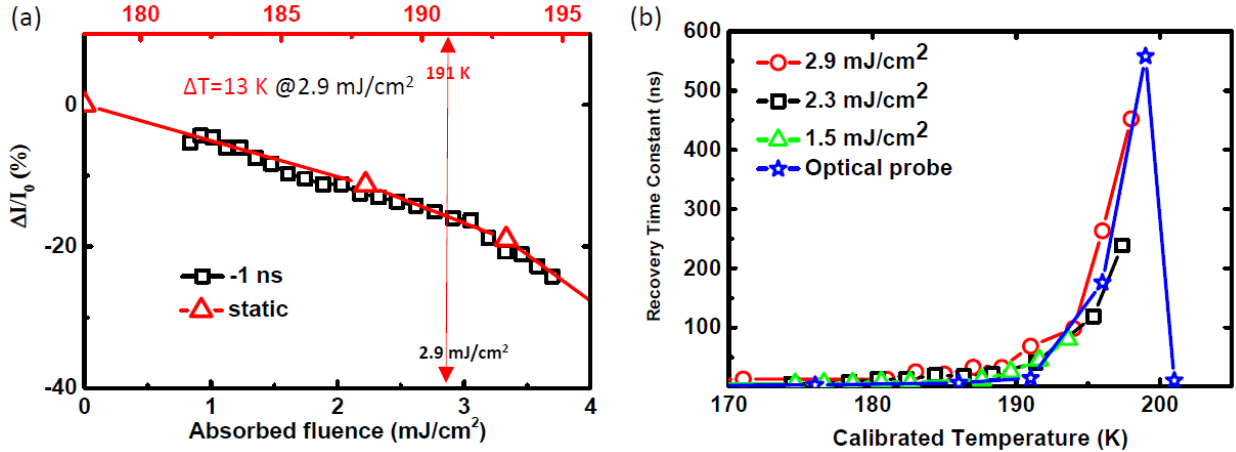

FIG. S6 Temperature calibration. (a) Relative CO peak intensity change as a function of the absorbed pump laser fluence. The sample temperature is 178 K (measured without laser illumination). Red curve is the static temperature-dependent CO peak intensity change. (b) Slowing down of the CO peak recovery time constant measured at different absorbed pump laser fluences. Curves are from time-resolved x-ray diffraction probe measurements under different pump laser fluences. Optical pump-probe measurements from Fig. 2b are also shown for comparison.

## 6. Fitting procedures

The time-dependent curves, such as those shown in Fig. 3(d) and 3(d) were fit by an exponential delay function: $I(t) = I_0 \exp\left(-\frac{t}{\tau}\right) + c$, where $I_0$ and $c$ are constants and $\tau$ is the characteristic decay time.

The temperature-dependent recovery time constant of CO diffraction intensity (Fig. 3e) is fit by the scaling law: $\tau = \tau_0 (1-T/T_c)^{-\Delta}$. By fixing $T_c = 200$ K as experimentally measured, the best fit yields $\tau_0 = 1.5 \pm 0.6$ ns and $\Delta = 1.25 \pm 0.10$. Since the relative difference of the sample temperature with respect to $T_c$ is critical for the fitting results, we studied the impact of the temperature uncertainty of 1 K on the fitting results. Using x-ray data as an example, the fitting results at various $T_c$ are summarized in Table S1. Considering the temperature uncertainty of 1 K in the experiment, we show the critical exponent as the mean of three $\Delta$ at three $T_c$ and the error bar as the standard error of the mean: $\Delta = 1.24 \pm 0.19$. The increase of error bars does not affect the conclusions of the paper.

| Fitting results | $T_c$ (K) | | |
|---|---|---|---|
|  | 199 | 200 | 201 |
| $\Delta$ (x-ray) | 0.90 ± 0.09 | 1.25 ± 0.10 | 1.57 ± 0.11 |
| R-square | 0.947 | 0.968 | 0.977 |

Table S1. The fitting results at various fixed $T_c$

To compare with the critical scaling of the 2D Ising model, we fixed $\Delta = 2.16$ [1,2] and $T_c = 200$ K, but the best-fit curve deviates from data points (Fig. S7). To compare with the nucleation and growth model, the time constant of the recovery of CO diffraction intensity $\tau$ is proportional to the temperature-dependent nucleation rate of CO domains $N = N_0 \exp(-\frac{B}{(T-T_c)^2})$ [3]. Therefore, $\tau = \tau_0 \exp\left(\frac{B}{(T-T_c)^2}\right)$. Since $\tau_0 \sim 3$ ns as shown in our measurements, we fixed $\tau_0 = 3$ ns and $T_c = 200$ K, the fit yields B=20 K$^2$ and is shown in the dotted line of Fig. S7. The deviation of the dotted curve from the measured data points indicates that the temperature-dependent nucleation and growth does not explain the observed slowing down.

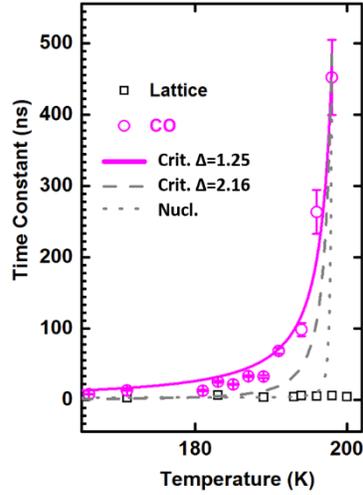

FIG. S7 Scaling of the slowing down. The magenta and dashed gray curves show the fit to power law with exponent $\Delta$ = 1.25 and 2.16 respectively. The dotted curve shows the fit to the nucleation and growth model.